\documentclass[12pt]{article}
\pdfoutput=1
\usepackage{jheppub}
\usepackage{showlabels}
\usepackage[utf8]{inputenc}
\usepackage{verbatim}
\usepackage{amsmath}
\usepackage{amssymb}
\usepackage{amsthm}
\usepackage{slashed}
\usepackage{amsfonts}

\newcommand{\be}{\begin{equation}}
\newcommand{\ee}{\end{equation}}
\newcommand{\bfig}{\begin{figure}\begin{center}}
\newcommand{\efig}{\end{center}\end{figure}}
\newcommand{\bi}{\begin{itemize}}
\newcommand{\ei}{\end{itemize}}

\newcommand{\lan}{\langle}
\newcommand{\ran}{\rangle}
\newcommand{\Tr}{\mathrm{Tr}}

\newtheorem{conj}{Conjecture}
\theoremstyle{definition}

\begin{document}
 \begin{flushright}
	\hfill{\tt MIT-CTP 5323, CALT-TH 2021-032, IPMU 21-0055}
		\end{flushright}
\title{A universal formula for the density of states in theories with finite-group symmetry}
\author[a]{Daniel Harlow}
\author[b,c]{and Hirosi Ooguri}
\affiliation[a]{Center for Theoretical Physics\\ Massachusetts Institute of Technology, Cambridge, MA 02139, USA}
\affiliation[b]{Walter Burke Institute for Theoretical Physics\\ California Institute of Technology,  Pasadena, CA 91125, USA}
\affiliation[c]{Kavli Institute for the Physics and Mathematics of the Universe (WPI)\\ University of Tokyo,
   Kashiwa, 277-8583, Japan}
\emailAdd{harlow@mit.edu, ooguri@caltech.edu}
\abstract{In this paper we use Euclidean gravity to derive a simple formula for the density of black hole microstates which transform in each irreducible representation of any finite gauge group.  Since each representation appears with nonzero density, this gives a new proof of the completeness hypothesis for finite gauge fields.  Inspired by the generality of the argument we further propose that the formula applies at high energy in any quantum field theory with a finite-group global symmetry, and give some evidence for this conjecture.}
\maketitle

\section{Introduction}
Black holes which are charged under a finite gauge group have so far received substantially less attention than their brethren with continuous charge.  Indeed most work which has been done has insisted on embedding the finite gauge group into a continuous one using the Higgs mechanism \cite{Krauss:1988zc,Alford:1989ch,Coleman:1991ku,GarciaGarcia:2018tua}, which leads to various inessential complications.  In this paper we use Euclidean gravity coupled directly to a finite gauge field to study the following question: in a quantum gravity theory with a long-range gauge symmetry with finite gauge group $G$, at fixed energy  how many black hole microstates transform in each irreducible representation of $G$?  This is a natural question in black hole physics, which as far as we know has not been studied outside of a few particular examples in string theory \cite{Sen:2009md,Sen:2010ts,Mandal:2010cj}.  In this paper we give the following general answer: in the semiclassical regime the density $\rho_\alpha(E)$ of black hole microstates transforming in each irreducible representation $\alpha$ of $G$ is given by
\be
\rho_{\alpha}(E)=\frac{d_\alpha^2}{|G|}\rho(E)\label{rhointro},
\ee
where $d_\alpha$ is the dimension of $\alpha$, $|G|$ is the number of elements of $G$, and 
\be
\rho(E)=\sum_\alpha \rho_\alpha(E)
\ee
is the total density of states.  We further propose that \eqref{rhointro} holds as a general statement about the high-energy density of states in any quantum field theory with a finite-group global symmetry, and we give some evidence for this conjecture.  

A few years ago a formula quite similar to \eqref{rhointro} appeared in the context of $BF$ gauge theory with \textit{continuous} gauge group $G$ coupled to Jackiw-Teitelboim gravity in $1+1$ dimensions \cite{Kapec:2019ecr}.  Indeed using results from \cite{Witten:1991we}, the authors of \cite{Kapec:2019ecr} showed that the disk partition function restricted to states in representation $\alpha$ is given by
\be\label{kmsform}
Z_\alpha(\beta)=\frac{d_\alpha^2}{\mathrm{Vol}(G)}e^{-\beta c_2(\alpha)/2}Z_{JT}(\beta),
\ee
where $\mathrm{Vol}(G)$ is the volume of $G$, $Z_{JT}(\beta)$ is the disk partition function of Jackiw-Teitelboim gravity, and $c_2(\alpha)$ is the quadratic Casimir invariant of $\alpha$.  Our formula \eqref{rhointro} can be thought of as a discrete version of this formula, but the topological ingredients underlying \eqref{kmsform} hold with much more generality when $G$ is finite so \eqref{rhointro} holds much more broadly than \eqref{kmsform}.

\section{Counting microstates with finite gauge charge}\label{finitesec}
\bfig
\includegraphics[height=5cm]{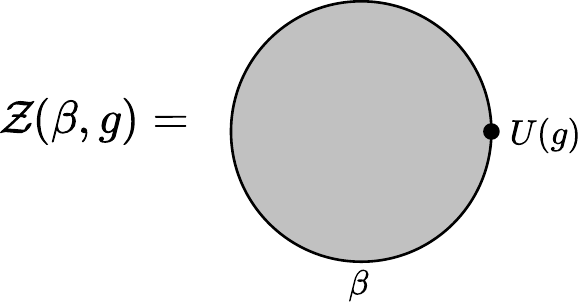}
\caption{The Euclidean calculation of the twisted black hole partition function: we fill in a boundary temporal circle of length $\beta$ with the Euclidean Schwarzschild geometry, but with an asymptotic symmetry operator $U(g)$ inserted.}\label{discreteZfig}
\efig
The main novelty of finite hair on black holes is that it does not contribute to the energy-momentum tensor, and thus does not change the geometry of the black hole.\footnote{If the finite gauge group is obtained from Higgsing a continuous group there is some backreaction right around the black hole \cite{Krauss:1988zc,Alford:1989ch,Coleman:1991ku,GarciaGarcia:2018tua}, but for big enough black holes this backreaction decays exponentially with distance so the only hair which is detectable far away is that which is described by the pure finite gauge theory we study here.}  In studying the microstates of a black hole which carries finite gauge charge, a natural object to consider is the twisted partition function
\be\label{Zfin}
\mathcal{Z}(\beta,g)\equiv \Tr\left(e^{-\beta H}U(g)\right).
\ee
Here $U(g)$ is the unitary operator at spatial infinity which implements the finite gauge symmetry.  In the continuous case $\mathcal{Z}(\beta,g)$ can be thought of as a grand canonical partition function with an imaginary chemical potential.  Computing $\mathcal{Z}(\beta,g)$ directly from \eqref{Zfin} requires detailed non-perturbative information about quantum gravity, but we can compute it without such knowledge by using the magic of Euclidean quantum gravity.  

\bfig
\includegraphics[height=5cm]{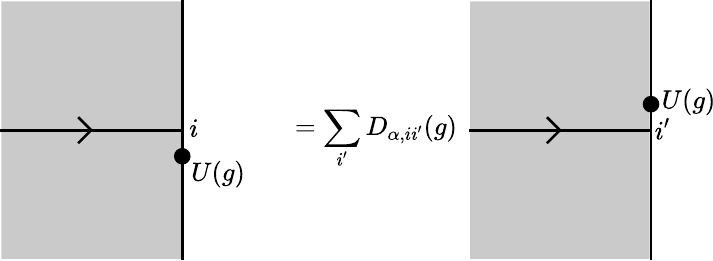}
\caption{The rule for moving an asymptotic symmetry operator past a Wilson line endpoint.}\label{wilsonfig}
\efig
\bfig
\includegraphics[height=8cm]{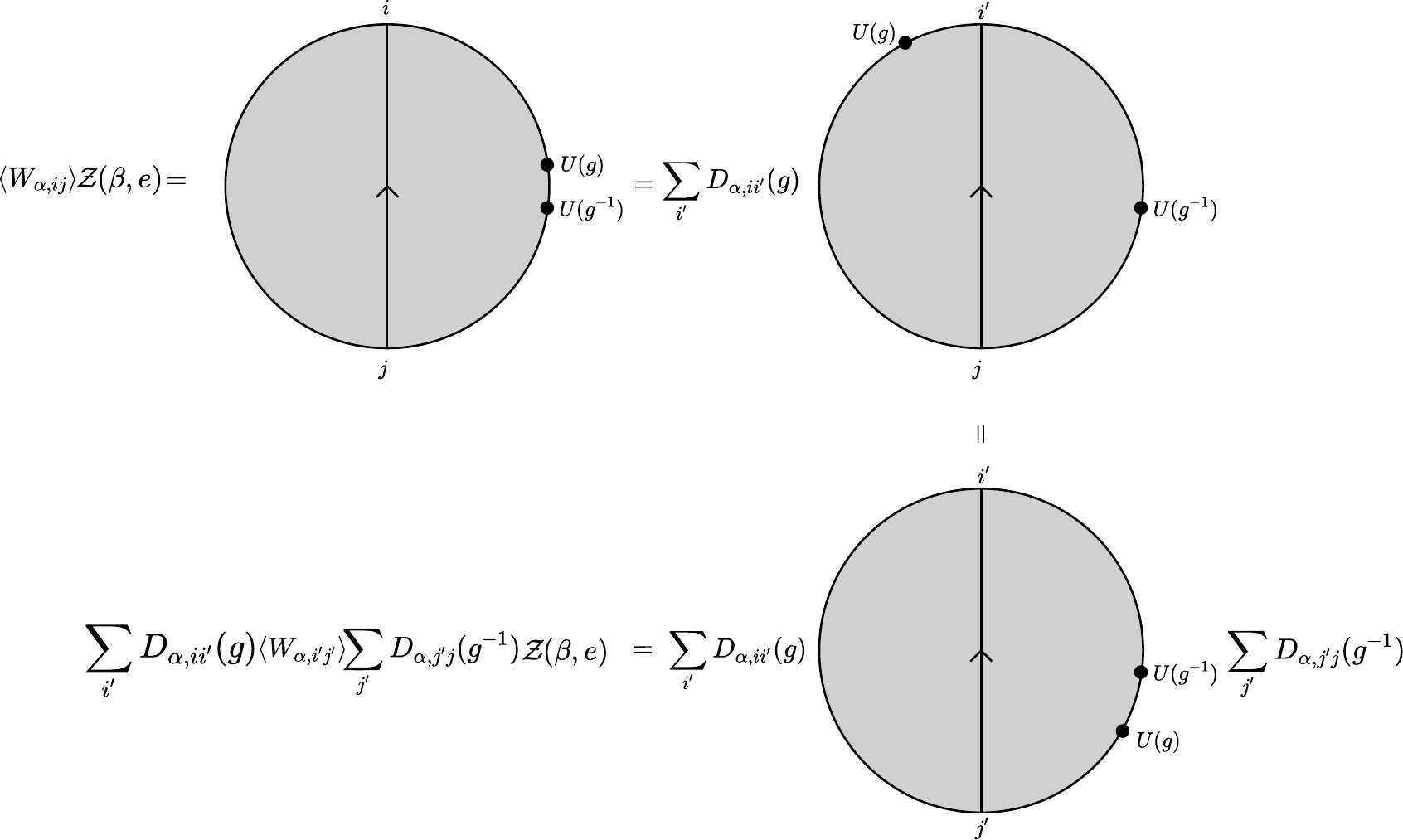}
\caption{Showing that $\lan W_{\alpha,ij}\ran=C_\alpha \delta _{ij}$.  By introducing a trivial factor of $U(g)U(g^{-1})$ and then moving $U(g)$ around the thermal circle using the rule from figure \ref{wilsonfig} (and its conjugate), we see that $\lan W_{\alpha,ij}\ran$ must be invariant under conjugation.  Since $\alpha$ is irreducible, by Schur's lemma $\lan W_{\alpha,ij}\ran$ must therefore be proportional to the identity.  We can interpret $\lan W_{\alpha,ij}\ran \mathcal{Z}(\beta,e)$ as the norm of the unnormalized thermofield double state with a background charge inserted, so $C_\alpha$ is positive semi-definite.  In fact it is strictly positive, as we are assuming that the gauge field is in a deconfined phase so this norm should be nonzero.}\label{wilsonproof}  
\efig
The Euclidean expression for $\mathcal{Z}(\beta,g)$ is illustrated in figure \ref{discreteZfig}.  We will evaluate it by inserting a Wilson line $W_{\alpha}$, where $\alpha$ indicates an irreducible representation of $G$, and then using the defining property of $U(g)$, which is that moving $U(g)$ across the endpoint of a Wilson line $W_{\alpha}$ multiplies $W_\alpha$ on the left by the representation matrix $D_\alpha(g)$ (see figure \ref{wilsonfig}).  For more on why this is the correct rule, see e.g. section 3 of \cite{Harlow:2018tng}.  In particular we can use this rule to show that the thermal expectation value of the Wilson line is proportional to the identity:
\be\label{Weq}
\lan W_{\alpha,ij}\ran=C_\alpha \delta_{ij},
\ee
with 
\be
C_\alpha>0.
\ee
We show the argument in figure \ref{wilsonproof}. 

\bfig
\includegraphics[height=8cm]{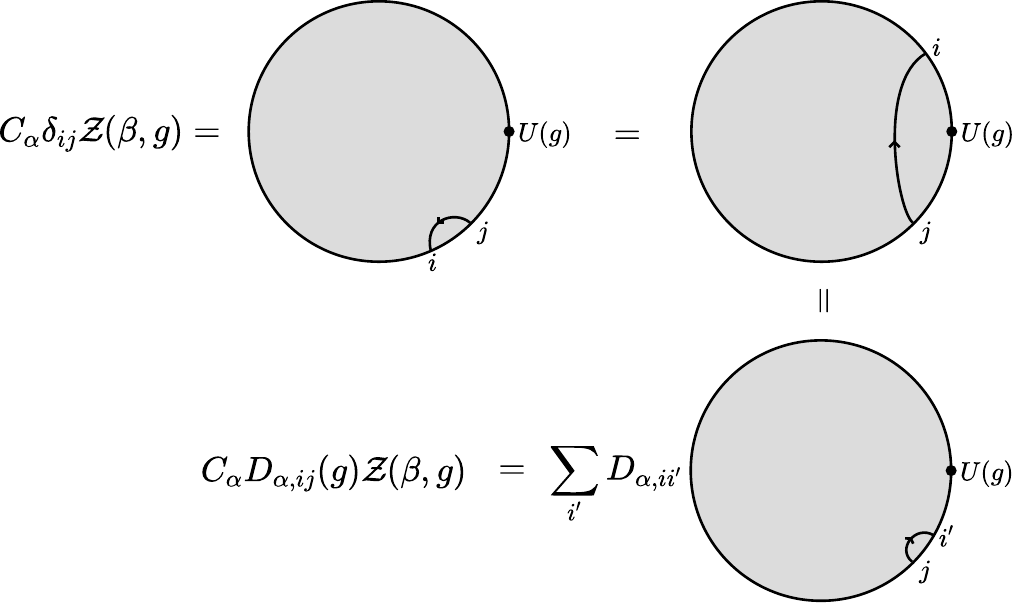}
\caption{Deriving \eqref{Zresult1}.  Since $G$ is a finite group the Wilson line $W_{\alpha,ij}$ is a topological operator, so we can move one endpoint around the thermal circle, picking up a group transformation along the way.  In the first and last steps we use cluster decomposition/locality: when the Wilson line is arbitrarily small and far from $U(g)$, we can replace it by its expectation value $\lan W_{\alpha,ij}\ran=C_\alpha \delta_{ij}$.}\label{Zcalcfig}
\efig
We now use these observations to compute $\mathcal{Z}(\beta,g)$.  We can do so by using the fact that the Wilson line is a topological operator, which can be freely deformed away from other operators in the path integral, to show that for any $\alpha$ we have
\be\label{Zresult1}
C_\alpha\delta_{ij}\mathcal{Z}(\beta,g)=C_\alpha D_{\alpha,ij}(g)\mathcal{Z}(\beta,g).
\ee
Since $C_\alpha>0$ for all $\alpha$, and for any $g\neq e$ we can always find some $\alpha$ such that $D_{\alpha,ij}(g)\neq \delta_{ij}$ (this is a consequence of the Peter-Weyl theorem, see e.g. appendix A of \cite{Harlow:2018tng}), we see that we must have
\be\label{Zresult2}
\mathcal{Z}(\beta,g)=Z(\beta)\delta(g),
\ee
where 
\be
\delta(g)\equiv \begin{cases} 1 & g=e\\ 0 & g\neq e\end{cases}
\ee
and
\be
Z(\beta)\equiv \mathcal{Z}(\beta,e)
\ee
is the usual thermal partition function.  The proof of \eqref{Zresult1} is shown in figure \ref{Zcalcfig}.  Note the key role of the contraction of the thermal circle in this argument: otherwise the Wilson line could not be deformed in this manner.  As emphasized in \cite{Harlow:2020bee},  this contraction lies at the heart of the ability of Euclidean quantum gravity to compute partition functions such as \eqref{Zfin} without knowing the full set of microstates, and allowing it is an assumption about the non-perturbative nature of quantum gravity.  In \cite{Harlow:2020bee} it was argued that the justification for this is ultimately holographic in nature.
   
Let's now see what \eqref{Zresult2} tells us about the spectrum of charged states.  We will assume that the spectrum of states is dense enough that to a good approximation we can compute the partition function using a smooth density of states $\rho(E)$,
\be
\mathcal{Z}(\beta,e)=\int_0^\infty dE\rho(E)e^{-\beta E},
\ee
and we will further assume that we can decompose this density of states by irreducible representation as
\be\label{rhosum}
\rho(E)=\sum_\alpha \rho_\alpha(E),  
\ee
where each $\rho_\alpha(E)$ is itself a smooth (and in fact probably analytic) function of $E$.\footnote{If these smoothness assumptions don't hold then $\mathcal{Z}(\beta,g)$ will not be a well-behaved function of $\beta$, which is contrary to our expectations from Euclidean gravity (or just generic quantum chaos).}  From \eqref{Zfin} we then have
\be
\mathcal{Z}(\beta,g)=\sum_\alpha\frac{1}{d_\alpha}\int_0^\infty dE \rho_\alpha(E)\chi_\alpha(g)e^{-\beta E},
\ee
where $\chi_\alpha(g)$ is the character
\be
\chi_\alpha(g)\equiv \Tr D_\alpha(g),  
\ee
and so from \eqref{Zresult2} we have
\be
\sum_\alpha\frac{1}{d_\alpha}\int_0^\infty dE \rho_\alpha(E)\chi_\alpha(g)e^{-\beta E}=\delta(g)\int_0^\infty dE\rho(E)e^{-\beta E}.
\ee
Multiplying each side of this equation by $\chi^*_{\alpha'}(g)$ and then averaging over the group element $g$ using the Schur orthogonality relation (see e.g. appendix A of \cite{Harlow:2018tng})
\be
\frac{1}{|G|}\sum_g\chi^*_{\alpha'}(g)\chi_\alpha(g)=\delta_{\alpha\alpha'},
\ee
we have
\begin{align}\nonumber
\int_0^\infty dE\frac{\rho_{\alpha'}(E)}{d_{\alpha'}}e^{-\beta E}&=\int_0^\infty dE \frac{\chi^*_{\alpha'}(e)}{|G|}\rho(E) e^{-\beta E}\\
&=\int_0^\infty dE \frac{d_{\alpha'}}{|G|}\rho(E)e^{-\beta E},
\end{align}
so taking the inverse Laplace transform we at last find
\be\label{rhoresult}
\rho_\alpha(E)=\frac{d_\alpha^2}{|G|}\rho(E).
\ee
We emphasize that \eqref{rhoresult} does not say that all representations appear with equal multiplicities, it instead says that higher-dimensional representations are favored by a factor of $d_\alpha$.  We can check that \eqref{rhosum} indeed holds, which follows from the usual relation
\be
\sum_\alpha d_\alpha^2=|G|,
\ee
which is again a consequence of Schur orthogonality and the Peter-Weyl theorem. 

It may be helpful to illustrate the connection between \eqref{Zresult2} and \eqref{rhoresult} more concretely.  Indeed say we have $N$ states transforming in representations of $\mathbb{Z}_2$, with $\alpha N$ in the trivial representation and $(1-\alpha)N$ in the sign representation.  We can think of $N$ as being very large.  The twisted partition function is given by
\be
\mathcal{Z}(g)=\begin{cases}
N & g=1\\
(2\alpha-1)N & g=-1
\end{cases}.
\ee
This is maximized at the identity for any value of $\alpha$, but the ratio of $Z(1)$ to $Z(-1)$ will not be large unless $\alpha$ is very close to $1/2$: we indeed need the states to obey \eqref{rhoresult} in order for \eqref{Zresult2} to hold.  Similarly if we instead have $N$ states transforming in representations of $\mathbb{Z}_3$, with $\alpha_1 N$ states in the defining representation, $\alpha_2 N$ states in its complex conjugate, and $(1-\alpha_1-\alpha_2)N$ states in the trivial representation, then we have
\be
\mathcal{Z}(g)=\begin{cases}
N & g=1\\
(\alpha_1 e^{2\pi i/3}+\alpha_2 e^{-2\pi i/3}+1-\alpha_1-\alpha_2)N & g=e^{2\pi i /3}\\
(\alpha_1 e^{-2\pi i/3}+\alpha_2 e^{2\pi i/3}+1-\alpha_1-\alpha_2)N & g=e^{-2\pi i /3}
\end{cases}.
\ee
The second two of these will both be much smaller than the first if and only if $\alpha_1\approx \alpha_2\approx 1/3$, again as predicted by \eqref{rhoresult}. These examples hopefully make it clear that our formula \eqref{rhoresult} is not merely a consequence of generic phase cancellations in the twisted partition function when $g\neq e$. 

We briefly comment on what kinds of corrections exist to \eqref{Zresult2}.  The topological nature of the argument ensures there are no perturbative corrections, but non-perturbative corrections are definitely present.  Perhaps the most interesting arise from branes with $(d-2)$-dimensional worldvolume around which the gauge field has non-vanishing holonomy \cite{Krauss:1988zc,GarciaGarcia:2018tua}, which should always exist in a gravity theory with a finite group gauge symmetry \cite{Heidenreich:2021tna}.  We can insert such a brane at the Euclidean horizon in figure \ref{discreteZfig}, which gives a contribution to $Z(\beta,g)$ which need not vanish when $g\neq e$.  Such contributions are suppressed however by a factor of $e^{-AT}$, where $A$ is the horizon area and $T$ is the tension of the brane, so they give only exponentially small corrections to \eqref{Zresult2} in the semiclassical limit.

\section{A general conjecture}\label{cmtsec}

The argument which led to \eqref{rhoresult} used almost no features of the quantum gravity theory: we did not have to use the Einstein action or assume anything about what matter fields might be present, and we could even have changed the boundary spatial manifold.  The only real assumption is that we are in a regime where the dominant saddle has the property that the thermal circle contracts to zero size, and at least in $AdS_d$ this should always be true on any compact boundary manifold for sufficiently high energy.  Inspired by this generality we propose the following conjecture:

\begin{conj}\label{conjecture1}
In any quantum field theory with a finite-group global symmetry $G$, on any compact spatial manifold at sufficiently high energy the density of states  $\rho_\alpha(E)$ for each irreducible representation $\alpha$ of $G$ obeys \eqref{rhointro}.
\end{conj}

\bfig
\includegraphics[height=6cm]{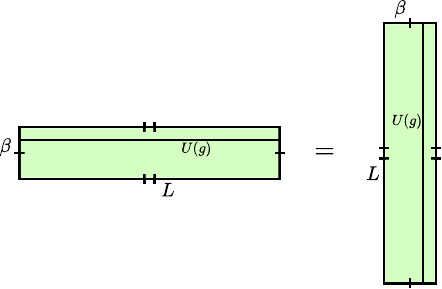}
\caption{Proving conjecture \ref{conjecture1} for $1+1$ dimensional CFTs.  By modular invariance, the high-temperature limit of $\mathcal{Z}(\beta,g)$ is the same as the low-temperature limit of the thermal trace in a sector twisted by $g$.}\label{cardyfig}
\efig
The first evidence for this conjecture is that it was recently shown to be true in any conformal field theory in $1+1$ dimensions, and thus also any $1+1$ dimensional quantum field theory which flows from a conformal field theory at short distances \cite{Pal:2020wwd}.  In $1+1$ dimensions the only compact spatial manifold is the circle, which we take to have circumference $L$. The idea is to use the same modularity argument as goes into deriving the Cardy formula \cite{Cardy:1986ie,Pal:2020wwd}: the partition function $\mathcal{Z}(\beta,g)$ defined by \eqref{Zfin} at high temperature is related to a low-temperature thermal partition function in a sector which is twisted by $g$ as we go around the spatial circle (see figure \ref{cardyfig}).  The latter limit is dominated by its ground state, which if we gauge the $G$ symmetry leads to twisted operator of scaling dimension $\Delta_g$, and the energy of this ground state is given by
\be
E_g=\frac{2\pi}{\beta}\left(\Delta_g-\frac{c}{12}\right),
\ee
where the constant shift is the usual Casimir energy which comes from the Schwarzian derivative \cite{Cardy:1986ie}.  Therefore at high temperature we have
\be
\mathcal{Z}(\beta,g)\approx e^{\frac{2\pi L}{\beta}\left(\frac{c}{12}-\Delta_g\right)}.
\ee
In particular if we set $g=e$ then the ``twisted'' ground state is just the vacuum, which corresponds to the identity operator with $\Delta_e=0$, and we recover the usual Cardy formula
\be
Z(\beta)=\mathcal{Z}(\beta,e)\approx e^{\frac{\pi Lc}{6\beta}}.
\ee
More generally we have
\be\label{2dresult}
\mathcal{Z}(\beta,g)\approx Z(\beta)e^{-\frac{2\pi L}{\beta}\Delta_g}.
\ee
Finally we expect that we should have $\Delta_g>0$ for any $g\neq e$, as the twisted boundary conditions will introduce field gradients which cause the energy to increase (we do not know of a rigorous derivation of this statement for 
a general conformal field theory, but it is certainly true in all known examples and in fact typically $\Delta_g\sim c$ \cite{Vafa:1985nm,Dixon:1985jw,Gepner:1987vz,Dijkgraaf:1989hb,Greene:1990ud}), and thus at high temperature we have
\be\label{Zdelt}
\mathcal{Z}(\beta,g)\approx Z(\beta)\delta(g), 
\ee
which is precisely the same as \eqref{Zresult2} and thus implies \eqref{rhointro} by the same steps as before \cite{Pal:2020wwd}.  The fact that we can obtain \eqref{Zresult2} both from this argument and also from the gravity argument of the previous section can be viewed as a check of the AdS/CFT correspondence with two boundary spacetime dimensions.\footnote{It is interesting to note that this derivation of \eqref{2dresult} also works if $G$ is a continuous group, and there we can explicitly confirm that $\Delta_g>0$ for $g\neq e$ thanks to the expression of the Noether
currents in terms of free bosons and parafermions \cite{Gepner:1987sm}.  In $AdS_3$ long-range bulk gauge theories with continuous gauge group inevitably include Chern-Simons terms, which make the gauge field topological in the infrared and thus allow the derivation of the previous section to go through also for a continuous gauge group.  In higher dimensions a gauge field with a continuous gauge group is typically not topological, but if we work in the limit of fixed charge and large energy then the background flux is small so the Wilson line can still be smoothly deformed at least at leading order in perturbation theory: a somewhat less robust version of \eqref{rhointro} thus still holds.}  

We can also give a heuristic combinatoric argument for \eqref{rhointro}.\footnote{This argument is based on the proof of theorem A.11 in \cite{Harlow:2018tng}, which is originally due to \cite{Levy:2003my}.}  Namely let's imagine that we have some finite collection of ``fundamental'' operators transforming in a faithful representation $\rho$ of $G$, with the property that all states of the theory are obtained by acting repeatedly on the vacuum with these operators and their hermitian conjugates.  As a model of the set of states with some large fixed energy, we will take the set of states which we obtain by acting $2n$ times with either the identity, an element of $\rho$, or an element of $\rho^\dagger$.  The Reeh-Schlieder theorem tells us that no linear combination of these operators can annihilate the vacuum, so this gives a set of nonzero states which transform in the representation
\be
\rho_n\equiv \left(1\oplus \rho \oplus \rho^\dagger\right)^{2n}.  
\ee 
The character of this representation is
\be
\chi_{\rho_n}(g)=\big(1+2\mathrm{Re}(\chi_\rho(g))\big)^{2n},
\ee
so we can count the number of times some particular irreducible representation $\alpha$ appears in $\rho_n$ by
\be
N_\alpha^{\{n\}}=\frac{1}{|G|}\sum_g\chi_\alpha(g)\big(1+2\mathrm{Re}(\chi_\rho(g))\big)^{2n}.
\ee
Noting that 
\be
\big(1+2\mathrm{Re}(\chi_\rho(g))\big)^2\leq (1+2d_\rho)^2
\ee
with equality if and only if $g=e$, we see that at large $n$ we have
\be
\lim_{n\to\infty}\frac{\big(1+2\mathrm{Re}(\chi_\rho(g))\big)^{2n}}{\sum_g \big(1+2\mathrm{Re}(\chi_\rho(g))\big)^{2n}}=\delta(g).
\ee
Therefore at large $n$ the ratio of $N_\alpha^{\{n\}}$ to $N_0^{\{n\}}$, where $N_0^{\{n\}}$ is the number of times the trivial representation appears in $\rho_n$, is given by
\begin{align}\nonumber
\lim_{n\to\infty}\frac{N_\alpha^{\{n\}}}{N_0^{\{n\}}}&=\sum_g \chi_\alpha(g)\lim_{n\to\infty}\frac{\big(1+2\mathrm{Re}(\chi_\rho(g))\big)^{2n}}{\sum_g \big(1+2\mathrm{Re}(\chi_\rho(g))\big)^{2n}}\\\nonumber
&=\chi_\alpha(e)\\
&=d_\alpha,
\end{align}
so the multiplicity of each irreducible representation is indeed weighted by $d_\alpha$ just as predicted by \eqref{rhointro} (in \eqref{rhointro} we have $d_\alpha^2$ since there we are counting states instead of representations).

Our formula \eqref{rhointro} can also be checked microscopically in string theory using existing results about certain extremal black holes with $\mathcal{N}=4$ supersymmetry which are charged under a $\mathbb{Z}_N$ gauge symmetry \cite{Sen:2009md,Sen:2010ts,Mandal:2010cj}.  There the twisted partition function $\mathcal{Z}(\beta,g)$ with $g\neq e$ was shown to be exponentially smaller than $\mathcal{Z}(\beta,e)$, both from a gravity computation and also directly by computing the twisted index from the worldsheet description.  Thus these results are compatible with \eqref{Zresult2}, and therefore \eqref{rhointro}.  It would be interesting to test \eqref{rhointro} for more stringy black holes with discrete gauge charge, and also to test conjecture \ref{conjecture1} in more quantum field theories.  Perhaps it could even be tested experimentally in condensed matter systems.  

\paragraph{Acknowledgments} We thank Luca Iliesiu, Juan Maldacena, Subir Sachdev, and Joaquin Turiaci for pointing out an error in a discussion of the weak gravity conjecture which appeared in the first version of this paper, and we particularly thank Luca and Joaquin for many useful explanations.  We also thank Chris Akers, Ben Heidenreich, Gary Horowitz, Henry Lin, Hong Liu, Juan Maldacena, Don Marolf, Sridip Pal, Matt Reece, Tom Rudelius, Ashoke Sen, and Cumrun Vafa for useful comments and discussion.  DH is supported by the Simons Foundation as a member of the ``It from Qubit'' collaboration, the Sloan Foundation as a Sloan Fellow, the Packard Foundation as a Packard Fellow, the Air Force Office of Scientific Research under the award number FA9550-19-1-0360, and the US Department of Energy under the task C grant DE-SC0012567. 
The work of HO is supported in part by the US Department of Energy, Office of Science, Office of High Energy Physics, 
under the award number DE-SC0011632, by the World Premier International Research Center Initiative, MEXT, 
Japan, and by JSPS Grant-in-Aid for Scientific Research 20K03965. This work was performed in part at Aspen Center for Physics, 
which is supported by National Science Foundation grant PHY-1607611.

\bibliographystyle{jhep}
\bibliography{bibliography}

\providecommand{\href}[2]{#2}\begingroup\raggedright\begin{thebibliography}{10}

\bibitem{Krauss:1988zc}
L.~M. Krauss and F.~Wilczek, {\it {Discrete Gauge Symmetry in Continuum
  Theories}},  {\em Phys. Rev. Lett.} {\bf 62} (1989) 1221.

\bibitem{Alford:1989ch}
M.~G. Alford, J.~March-Russell, and F.~Wilczek, {\it {Discrete Quantum Hair on
  Black Holes and the Nonabelian {Aharonov-Bohm} Effect}},  {\em Nucl. Phys. B}
  {\bf 337} (1990) 695--708.

\bibitem{Coleman:1991ku}
S.~R. Coleman, J.~Preskill, and F.~Wilczek, {\it {Quantum hair on black
  holes}},  {\em Nucl. Phys. B} {\bf 378} (1992) 175--246,
  [\href{http://arxiv.org/abs/hep-th/9201059}{{\tt hep-th/9201059}}].

\bibitem{GarciaGarcia:2018tua}
I.~Garcia~Garcia, {\it {Properties of Discrete Black Hole Hair}},  {\em JHEP}
  {\bf 02} (2019) 117, [\href{http://arxiv.org/abs/1809.03527}{{\tt
  arXiv:1809.03527}}].

\bibitem{Sen:2009md}
A.~Sen, {\it {A Twist in the Dyon Partition Function}},  {\em JHEP} {\bf 05}
  (2010) 028, [\href{http://arxiv.org/abs/0911.1563}{{\tt arXiv:0911.1563}}].

\bibitem{Sen:2010ts}
A.~Sen, {\it {Discrete Information from CHL Black Holes}},  {\em JHEP} {\bf 11}
  (2010) 138, [\href{http://arxiv.org/abs/1002.3857}{{\tt arXiv:1002.3857}}].

\bibitem{Mandal:2010cj}
I.~Mandal and A.~Sen, {\it {Black Hole Microstate Counting and its Macroscopic
  Counterpart}},  {\em Class. Quant. Grav.} {\bf 27} (2010) 214003,
  [\href{http://arxiv.org/abs/1008.3801}{{\tt arXiv:1008.3801}}].

\bibitem{Kapec:2019ecr}
D.~Kapec, R.~Mahajan, and D.~Stanford, {\it {Matrix ensembles with global
  symmetries and \textquoteright{}t Hooft anomalies from 2d gauge theory}},
  {\em JHEP} {\bf 04} (2020) 186, [\href{http://arxiv.org/abs/1912.12285}{{\tt
  arXiv:1912.12285}}].

\bibitem{Witten:1991we}
E.~Witten, {\it {On quantum gauge theories in two-dimensions}},  {\em Commun.
  Math. Phys.} {\bf 141} (1991) 153--209.

\bibitem{Harlow:2018tng}
D.~Harlow and H.~Ooguri, {\it {Symmetries in quantum field theory and quantum
  gravity}},  {\em Commun. Math. Phys.} {\bf 383} (2021), no.~3 1669--1804,
  [\href{http://arxiv.org/abs/1810.05338}{{\tt arXiv:1810.05338}}].

\bibitem{Harlow:2020bee}
D.~Harlow and E.~Shaghoulian, {\it {Global symmetry, Euclidean gravity, and the
  black hole information problem}},  {\em JHEP} {\bf 04} (2021) 175,
  [\href{http://arxiv.org/abs/2010.10539}{{\tt arXiv:2010.10539}}].

\bibitem{Heidenreich:2021tna}
B.~Heidenreich, J.~Mcnamara, M.~Montero, M.~Reece, T.~Rudelius, and
  I.~Valenzuela, {\it {Non-Invertible Global Symmetries and Completeness of the
  Spectrum}},  \href{http://arxiv.org/abs/2104.07036}{{\tt arXiv:2104.07036}}.

\bibitem{Pal:2020wwd}
S.~Pal and Z.~Sun, {\it {High Energy Modular Bootstrap, Global Symmetries and
  Defects}},  {\em JHEP} {\bf 08} (2020) 064,
  [\href{http://arxiv.org/abs/2004.12557}{{\tt arXiv:2004.12557}}].

\bibitem{Cardy:1986ie}
J.~L. Cardy, {\it {Operator Content of Two-Dimensional Conformally Invariant
  Theories}},  {\em Nucl. Phys. B} {\bf 270} (1986) 186--204.

\bibitem{Vafa:1985nm}
C.~Vafa and E.~Witten, {\it {Bosonic String Algebras}},  {\em Phys. Lett. B}
  {\bf 159} (1985) 265--268.

\bibitem{Dixon:1985jw}
L.~J. Dixon, J.~A. Harvey, C.~Vafa, and E.~Witten, {\it {Strings on
  Orbifolds}},  {\em Nucl. Phys. B} {\bf 261} (1985) 678--686.

\bibitem{Gepner:1987vz}
D.~Gepner, {\it {Exactly Solvable String Compactifications on Manifolds of
  SU(N) Holonomy}},  {\em Phys. Lett. B} {\bf 199} (1987) 380--388.

\bibitem{Dijkgraaf:1989hb}
R.~Dijkgraaf, C.~Vafa, E.~P. Verlinde, and H.~L. Verlinde, {\it {The Operator
  Algebra of Orbifold Models}},  {\em Commun. Math. Phys.} {\bf 123} (1989)
  485.

\bibitem{Greene:1990ud}
B.~R. Greene and M.~R. Plesser, {\it {Duality in Calabi-Yau Moduli Space}},
  {\em Nucl. Phys. B} {\bf 338} (1990) 15--37.

\bibitem{Gepner:1987sm}
D.~Gepner, {\it {New Conformal Field Theories Associated with Lie Algebras and
  their Partition Functions}},  {\em Nucl. Phys. B} {\bf 290} (1987) 10--24.

\bibitem{Levy:2003my}
T.~Levy, {\it {Wilson loops in the light of spin networks}},  {\em J. Geom.
  Phys.} {\bf 52} (2004) 382--397,
  [\href{http://arxiv.org/abs/math-ph/0306059}{{\tt math-ph/0306059}}].

\end{thebibliography}\endgroup
\end{document}